\documentclass[apj]{emulateapj}
\usepackage{apjfonts} 

\usepackage{graphicx}

\newcommand{\Fermi}{{\it{}Fermi}\ }

\makeindex
\citeindextrue

\received{July 30, 2009}
\revised{October 24, 2009}
\accepted{November 6, 2009}
\journalinfo{Astrophysical~journal Letters}
\submitted{}

\shorttitle{VLBI identification of the bright \Fermi LAT sources}
\shortauthors{Kovalev}

\begin{document}

\title{Identification of the Early \textit{Fermi} LAT Gamma-Ray Bright Objects with Extragalactic VLBI sources}

\author{Y. Y. Kovalev}
\affil{Max-Planck-Institut f\"ur Radioastronomie,
       Auf dem H\"ugel 69, 53121 Bonn, Germany;}
\affil{Astro Space Center of Lebedev Physical Institute,
       Profsoyuznaya 84/32, 117997 Moscow, Russia}
\email{yyk@asc.rssi.ru}

\begin{abstract}
A list of 205 $\gamma$-ray strong objects was reported recently as a
result of a 3-month integration with the Large Area Telescope on~board
the {\it Fermi Gamma-Ray Space Telescope}. We attempted identification
of these objects, cross-correlating the $\gamma$-ray positions with VLBI
positions of a large all-sky sample of extragalactic radio sources
selected on the basis of their parsec-scale flux density. The original
associations reported by the \Fermi team are confirmed and six new
identifications are suggested. A Monte-Carlo analysis shows that the
fraction of chance associations in our analysis is less than 5~per~cent,
and confirms that the vast majority of $\gamma$-ray bright extragalactic
sources are radio loud blazars with strong parsec-scale jets. A
correlation between the parsec-scale radio and $\gamma$-ray flux is
supported by our analysis of a complete VLBI flux-density-limited sample
of extragalactic jets. 
The effectiveness of using a VLBI catalog to find associations between
$\gamma$-ray detections and compact extragalactic radio sources,
especially near the Galactic plane, is demonstrated. It is suggested
that VLBI catalogs should be used for future identification of {\it
Fermi} LAT objects.

\end{abstract}


\section{Introduction}
\label{s:introduction}

The \textit{Fermi Gamma-Ray Space Telescope} (previously known as \textit{GLAST})
was successfully launched in June~2008. Even with the modern Large Area
Telescope on board \citep[LAT,][]{LAT09}, the positional uncertainty of
$\gamma$-ray measurements in the energy range from about 20~MeV to more
than 300~GeV remains relatively poor, typically $3\arcmin$ to $20\arcmin$.
This provides a challenge for $\gamma$-ray source
identification. A large fraction of the high energy $\gamma$-ray sources
detected by the EGRET telescope on board the {\it Compton Gamma Ray
Observatory} \citep{3EG} was identified with blazars
\cite[e.g.,][]{Mattox_etal01,SRM03,SRM04}. This fact critically helps in
identifying objects detected by the \Fermi LAT. In
order to help the identification process, several samples of blazars and blazar
candidates were constructed recently \citep{CRATES,CGRABS,BZCAT}. These
compilations of many thousands of extragalactic objects, together with
other catalogs covering radio to $\gamma$-ray bands, were successfully used
for identifying the LAT-detected sources by \cite{LATBSL}.

Results of parsec-scale Very Long Baseline Interferometry (VLBI)
measurements were not applied by \cite{LATBSL} in the process of
identifying bright LAT sources. However, this can provide important
extra information and improve the estimation of the probability of correct
identification since VLBI filters out
objects which do not host strong compact jets at parsec scales. The
latter is a strong characteristic of a radio loud blazar. VLBI can
especially help in the region around the Galactic plane where the available
multi-band coverage of the sky is much poorer due to Galactic
absorption.
An obvious weakness of the VLBI approach is the fact that radio
weak objects (e.g., high energy selected BL~Lacs) are missed in the
currently available large VLBI all~sky surveys.

During the last several decades, a number of large VLBI surveys were
conducted covering the frequency range between 2 and 100~GHz
\cite[e.g.,]{vcs1,VLBApls,PTZ03,2cmPaperIV,Ojha_etal04,Lovell_etal04,Hel07,Lanyi_etal05,sslee08}
including dedicated surveys in the Galactic plane
\citep[e.g.,][]{Petrov_etal07}. The largest in terms of the number of
sources observed is a VLBI effort at 2 and 8~GHz with a prime goal to
construct an extragalactic celestial reference frame
\citep{icrf98,icrf-ext2-2004} and to find extragalactic sources which
are compact at parsec scales and are suitable for phase referencing
\citep[][and references therein]{vcs6}. We note that while the majority
of those are of a blazar type, \cite{Hel07} have found 10~per~cent of
compact extragalactic radio sources to be optically associated with
galaxies from the Sloan Digital Sky Survey.
In this paper we use the 2/8~GHz VLBI surveys to associate the early
\Fermi LAT bright source list with radio counterparts at parsec scales,
make a statistical check on the reliability of these identifications,
and analyze the radio properties of the VLBI counterparts of the
$\gamma$-ray bright AGN.

\begin{figure*}[t]
\begin{center}
\resizebox{0.95\hsize}{!}{
  \includegraphics[trim=+0.0cm 0cm 0cm 0.0cm]{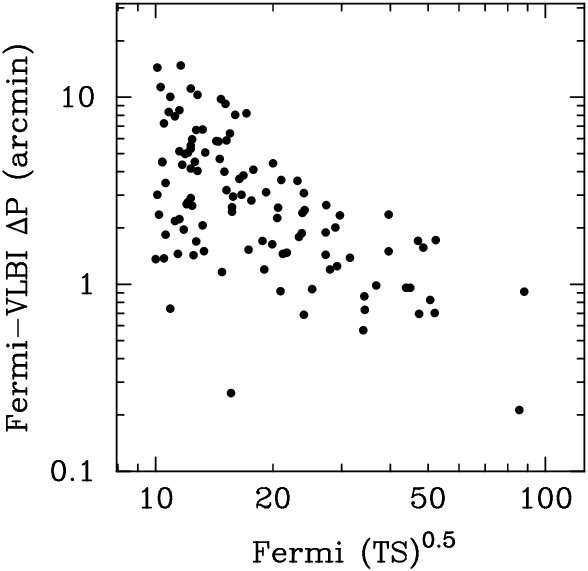}
  \includegraphics[trim=+0.7cm 0cm 0cm 0cm,clip]{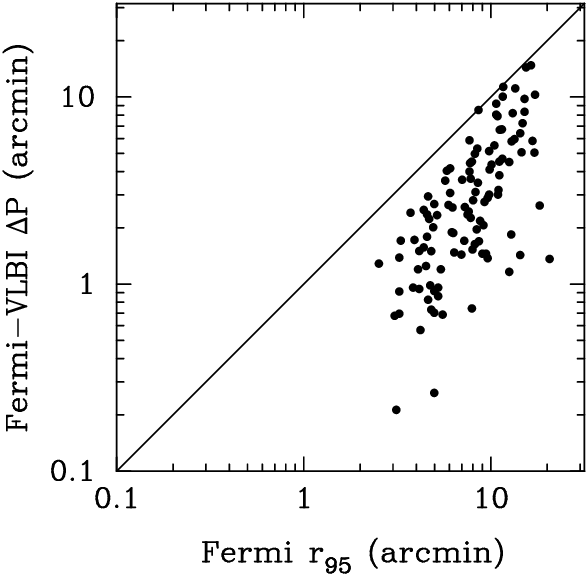}
  \includegraphics[trim=-0.3cm 0cm 0cm 0cm]{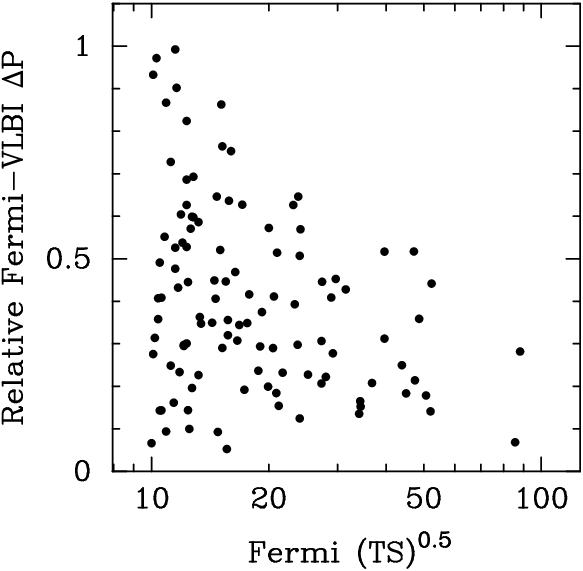}
}
\end{center}
\caption{\label{f:TSr95dP}
{\it Left panel:} Absolute difference $\Delta P$ between
\Fermi LAT positions and astrometric VLBI positions for positively
associated sources from Table~\ref{t:ass} versus
square root of \Fermi LAT Test Statistics from \cite{LATBSL}.
{\it Middle panel:} {\it Fermi}-VLBI $\Delta P$ difference 
versus \Fermi LAT 95~per~cent confidence radius from Abdo et al.
{\it Right panel:} Relative difference $\Delta P / r_{95}$
versus square root of \Fermi LAT Test Statistics.
}
\end{figure*}

\section{VLBI catalog in use}
\label{s:radio}

The VLBI positions for more than 4,000
extragalactic sources covering the whole sky were used in this
analysis\footnote{\url{http://astrogeo.org/vlbi/solutions/2009b\_astro/}}.
They are based on 2 and 8~GHz VLBI observations, most of which
were performed in the period 1994 to 2007
\citep{icrf98,icrf-ext2-2004,vcs1,vcs2,vcs3,vcs4,vcs5,vcs6,RDV09,LCS09}.
The accuracy of VLBI astrometric position determinations is typically better
than 1~milliarcsecond. For details of the observations, data processing and
analysis see the cited papers. \cite{vcs5} have made an additional
dedicated effort to construct a complete parsec-scale
flux-density-limited sample with the following characteristics: flat
radio spectrum with spectral index $\alpha>-0.5$ ($S\sim\nu^\alpha$),
parsec-scale 8~GHz flux density (integrated flux density over VLBI map)
$S_\mathrm{VLBI}>0.2$~Jy, declination $\delta>-30^\circ$, Galactic latitude
$|b|>1\fdg5$. Known Galactic objects were excluded. The full all-sky
astrometric 2/8~GHz VLBI source (AVS) catalog of 4338 objects is used for
cross-identification with the 3~month \Fermi LAT bright source (LBS) list
from \cite{LATBSL} while the complete flux-density-limited VLBI source
(FVS) sample of 1848 objects is used for statistical tests and
analysis.

\section{\Fermi LAT -- VLBI cross-identification}
\label{s:ass}

A number of authors have suggested a close connection between the bright
$\gamma$-ray and parsec-scale radio emission from analyzing EGRET
\cite[e.g.,][]{LV03} and early \Fermi \citep[e.g.,][]{FM2,FM1} data.
They concluded that the related $\gamma$-ray and parsec-scale radio
emission should originate in spatially close regions. This means that
VLBI positions might serve very well as an estimate of a location of the
source of $\gamma$-ray emission detected by \Fermi LAT for extragalactic
objects.
Taking this assumption as our starting point, we have applied a simple
approach to cross-identify bright \Fermi LAT sources from \cite{LATBSL}.
A \Fermi source will be called successfully identified with a VLBI
one if the difference between the \Fermi LAT position and 
the VLBI position of the AVS sample is less than the 95~per~cent
confidence radius, $r_{95}$, estimated by \cite{LATBSL}.

\subsection{Results of the cross-identification}
\label{s:res_ass}
 
\begin{figure}[b]
\begin{center}
\resizebox{1.01\hsize}{!}
{
  \includegraphics[trim=0cm 0cm 0cm 0cm]{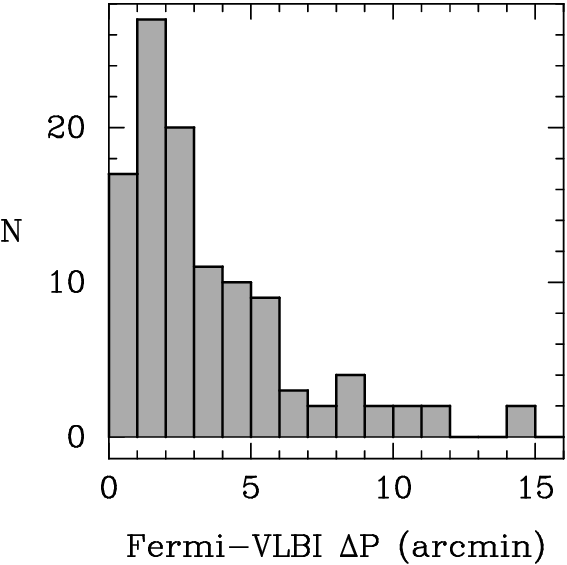}
  \includegraphics[trim=0.4cm 0cm 0cm 0cm,clip]{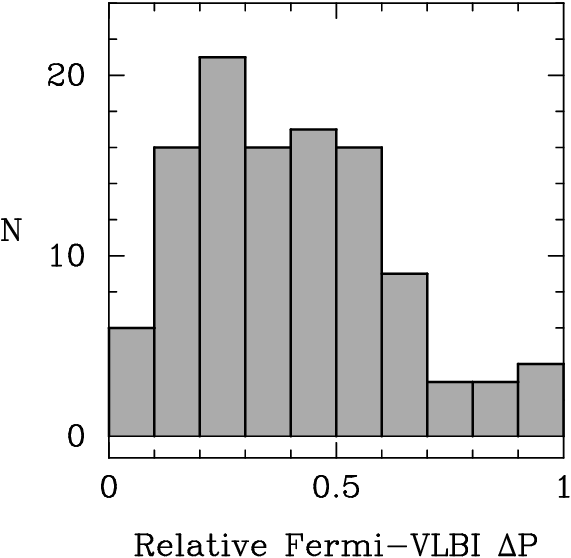}
}
\end{center}
\caption{\label{f:histdP}
Distribution of $\Delta P$ from Figure~\ref{f:TSr95dP}.
{\it Left panel:} Absolute difference $\Delta P$.
{\it Right panel:} Relative difference $\Delta P/r_{95}$.
}
\end{figure}

The results of this cross-identification can be found in
Table~\ref{t:ass}. We have successfully identified 111 out of 205 \Fermi
LAT sources. We note that the full list of the 205 bright $\gamma$-ray
sources presented by \cite{LATBSL} includes both Galactic and
extragalactic objects. No single \Fermi detection was identified with
more than one VLBI source. Figures~\ref{f:TSr95dP},\,\ref{f:histdP}
present the results of these identifications. The square root of the
\Fermi Test Statistics ($\sqrt{\mathrm{TS}}$) provides an estimate of
the signal-to-noise ratio (SNR) for every \Fermi LAT detection. As
expected, the difference, $\Delta P$, between the \Fermi position and
that of its VLBI identification becomes smaller with increasing
$\sqrt{\mathrm{TS}}$ (Figure~\ref{f:TSr95dP}). 
There are no identifications found with high \Fermi LAT SNR and high
relative \textit{Fermi}-VLBI positional difference, $\Delta P/r_{95}$.
This could be explained by an over-estimated systematic uncertainty which
was added in quadrature \citep{LATBSL} while calculating $r_{95}$. The
systematic uncertainty dominates the $r_{95}$ value for high-SNR \Fermi
sources becoming less dominant with decreasing \Fermi TS. Abdo et al.\
reported this systematic uncertainty to be conservative, they expect it
to improve.
The distribution of $\Delta P$ (Figure~\ref{f:histdP}) agrees well with
the results of \citet[Figure~9]{LATBSL}, who discuss this distribution
in detail.

Cross-identification of the \Fermi LBS versus AVS lists resulted in 111
identifications out of 4338 VLBI sources (2.6~per~cent of the VLBI
sample), while LBS versus FVS lists comparison gave 79 identifications out
of 1848 VLBI sources (4.3~per~cent). We note that the fraction of
identifications within the complete parsec-scale flux-density-limited
FVS list is about two times higher than that for the deeper AVS list of
sources. This could be the case if the bright $\gamma$-ray sources from
the LBS list ``prefer'' stronger parsec-scale radio counterparts, i.e.\
if there is a correlation between the $\gamma$-ray flux and parsec-scale
radio flux density. We note that the flux density cutoff of the FVS
sample is significantly higher than that for the AVS sample (200~mJy
versus about 50~mJy). The correlation suggested is indeed confirmed by
an analysis in \S~\ref{s:prop}.

\begin{deluxetable*}{llllllr}
\tablecolumns{7} 
\tabletypesize{\scriptsize} 
\tablewidth{0pt}  
\tablecaption{\label{t:ass}
VLBI associations of the \Fermi LAT bright sources
}
\tablehead{\colhead{LBS Name} & \colhead{VLBI name} &
\colhead{$\Delta P$} & \colhead{$r_{95}$} & \colhead{VLBI R.A.} &
\colhead{VLBI Dec.} & \colhead{$S_\mathrm{VLBI}$}\\
\colhead{(1)} & \colhead{(2)} &\colhead{(3)} & \colhead{(4)} & 
\colhead{(5)}  & \colhead{(6)} & \colhead{(7)}} 
\startdata 
0FGL J0643.2$+$0858 & 0640$+$090 & 0.043 & 0.121 & 06:43:26.4450 & $+$08:57:38.0126 & 0.61\\
0FGL J1123.0$-$6416 & 1121$-$640 & 0.039 & 0.125 & 11:23:19.4155 & $-$64:17:36.1970 & \nodata\\
0FGL J1328.8$-$5604 & 1325$-$558 & 0.058 & 0.142 & 13:29:01.1449 & $-$56:08:02.6657 & \nodata\\
0FGL J1604.0$-$4904 & 1600$-$489 & 0.037 & 0.078 & 16:03:50.6846 & $-$49:04:05.5050 & \nodata\\
0FGL J1830.3$+$0617 & 1827$+$062 & 0.067 & 0.097 & 18:30:05.9399 & $+$06:19:15.9521 & 0.41\\
0FGL J2001.0$+$4352 & 1959$+$437 & 0.025 & 0.069 & 20:01:12.8738 & $+$43:52:52.8391 & 0.07
\enddata 
\tablecomments{
Table~\ref{t:ass} is presented in its entirety in the electronic edition
of the Astrophysical Journal Letters. Here we present
six new suggested identifications for convenience and
guidance regarding Table's form and contents.
Columns are as follows:
(1) LAT bright source list catalog name \citep{LATBSL};
(2) IAU name (B1950.0) of VLBI association;
(3) Difference between \Fermi and VLBI position (deg);
(4) \Fermi LAT radius of 95~per~cent confidence region (deg) from \citep{LATBSL}
is shown for comparison with $\Delta P$;
(5) J2000.0 Right Ascension of VLBI association in hr, min, sec;
(6) J2000.0 Declination of VLBI association in angular hr, min, sec;
(7) Total parsec-scale 8~GHz flux density of VLBI association object (Jy).
}
\end{deluxetable*}

Comparison with the association results presented by \cite{LATBSL} has
shown the following. 105 \Fermi LAT detections have identifications common
to the Abdo et al.\ analysis and the current study. All the
extragalactic associations suggested by Abdo et al.\ but not found by
us, have radio flux densities less than 0.2~Jy or declinations
less than $-30^\circ$, i.e.\ belong to the flux
density and/or declination range where our VLBI catalog
is not complete. All the objects
which are identified by us but have no counterparts suggested by
Abdo et al.\ are located close to the Galactic plane --- a region of 
special difficulty for identification by the standard method of
Abdo et al., as discussed above.

\subsection{New identifications}
\label{s:new_ass}

The following six new identifications, not reported by \cite{LATBSL}, were
found by our analysis:
\object{0FGL~J0643.2+0858} (galactic latitude $b=2\fdg29$),
\object{0FGL~J1123.0-6416} ($b=-3\fdg02$),
\object{0FGL~J1328.8-5604} ($b=6\fdg41$),
\object{0FGL~J1604.0-4904} ($b=2\fdg54$),
\object{0FGL~J1830.3+0617} ($b=7\fdg54$),
\object{0FGL~J2001.0+4352} ($b=7\fdg12$).
All these are within $10^\circ$ from the Galactic plane.
Details on these sources can be found in Table~\ref{t:ass}.
We note that identifications with the same objects for two of these six
\Fermi detections were independently proposed recently by
\cite{Bassani_etal09} for \object{0FGL~J2001.0+4352} and \cite{MH09} for
\object{0FGL~J1830.3+0617} on the basis of a multi-band analysis. 
\cite{MH09} have also discussed possible flat-spectrum radio source
associations for \object{0FGL~J0643.2+0858} and
\object{0FGL~J1328.8$-$5604} which agree with our identifications.
\cite{Mirabal09} presented an X-ray point-source catalog from
\textit{Swift} XRT observations of unidentified \Fermi sources
including discussion of a potential counterpart for
\object{0FGL~J1604.0$-$4904}.

\subsection{Chance association analysis}
\label{s:MC}

We have performed a Monte-Carlo simulation to measure the chance
coincidence probability. We kept the \Fermi LBS catalog unchanged while
the VLBI catalog was scrambled; every source in the VLBI catalog was
shifted in a random position angle for a random angular distance on the
sky between $0^\circ$ and $5^\circ$. After that, cross-identification of
the catalogs was done as described above. This exercise was performed
1,000 times. The analysis was done independently for the LBS versus AVS
and LBS versus FVS lists. The mean number of chance associations was
found to be 4.5 (4.1~per~cent of 111 true identifications, see
\S\ref{s:res_ass}) and 2.9 (3.7~per~cent of 79 identifications) respectively.
This shows with a very high level of significance that
(\textit{i}) almost all of the associations found are firm identifications,
(\textit{ii}) the original assumption of the method that the
$\gamma$-ray sky is dominated by blazars with strong parsec-scale jets
is confirmed.

The described above method of VLBI catalog scrambling was applied to
represent the non-uniformity of the AVS list, where sky regions of a
special interest (e.g., the Galactic plane) are populated more densely.
The $5^\circ$-radius was chosen as an optimal size to represent the
non-uniformity being in the same time significantly greater than the
$r_{95}$ values.
We note that, for a uniform sky coverage in an original catalog, this
method reproduces uniformity in the Monte-Carlo test.
If we consider as a match \Fermi and VLBI sources with positional
difference less than $r_{95}/2$, we lose about 30~per~cent of
identifications (see Figures~\ref{f:TSr95dP},\,\ref{f:histdP}) while the
chance coincidence probability drops only down to about 2~per~cent.

\section{VLBI properties of the $\gamma$-ray identifications found}
\label{s:prop}

\begin{figure}[b]
\begin{center}
\resizebox{0.7\hsize}{!}
{
  \includegraphics[trim=0cm 0cm 0cm 0cm]{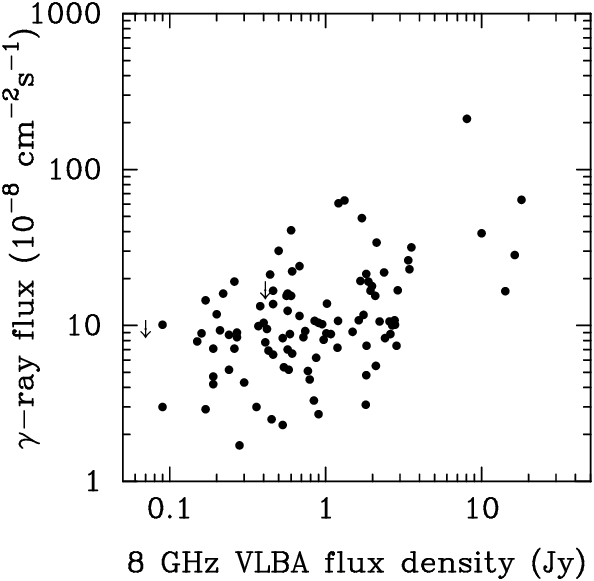}
}
\end{center}
\caption{\label{f:Sgr}
Average \Fermi LAT 100~MeV -- 1~GeV photon flux \citep{LATBSL} versus
8~GHz parsec-scale flux density non-simultaneously measured by VLBA.
The arrows represent an upper limit on the $\gamma$-ray flux.
}
\end{figure}

We have performed a correlation analysis between the average \Fermi LAT
100~MeV -- 1~GeV photon flux \citep{LATBSL} versus 8~GHz parsec-scale
flux density measured by VLBI between 1994 and 2008
(Figure~\ref{f:Sgr}). The non-parametric Kendall $\tau$ test confirms a
positive correlation in Figure~\ref{f:Sgr} at a confidence level greater
than $99.9$~per~cent for the 100~MeV -- 1~GeV energy band (the two upper
limits were ignored by the test which did not affect the conclusion).
This agrees with results of a similar analysis performed by \cite{FM2}
for simultaneous LAT $\gamma$-ray --- 15~GHz VLBA MOJAVE
\citep{Lister_etal09} measurements. Possible systematics in this dependence
resulting from the different properties of two populations of extragalactic
$\gamma$-ray sources (low energy peaked versus high energy peaked) were
mentioned by \cite{LBAS}. We do not go into detailed analysis of this in
the present paper since \cite{FM2} have addressed this issue in their
study.

\begin{figure}[t]
\begin{center}
\resizebox{0.7\hsize}{!}{
  \includegraphics[trim=0cm 0.8cm 0cm 0cm,clip]{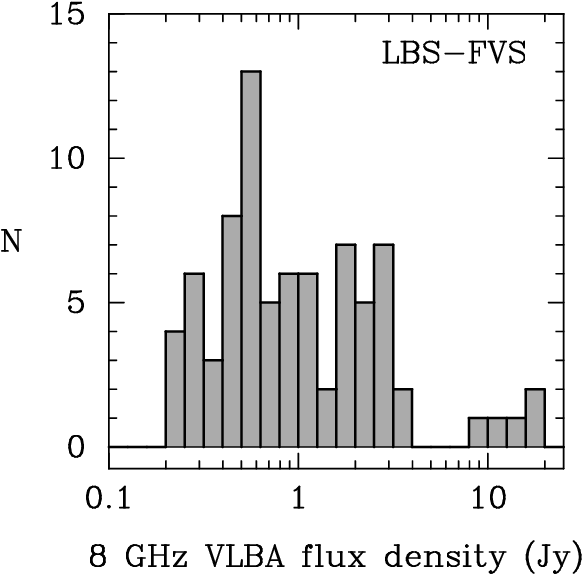}
}
\resizebox{0.7\hsize}{!}{
  \includegraphics[trim=0cm 0cm 0cm 0cm]{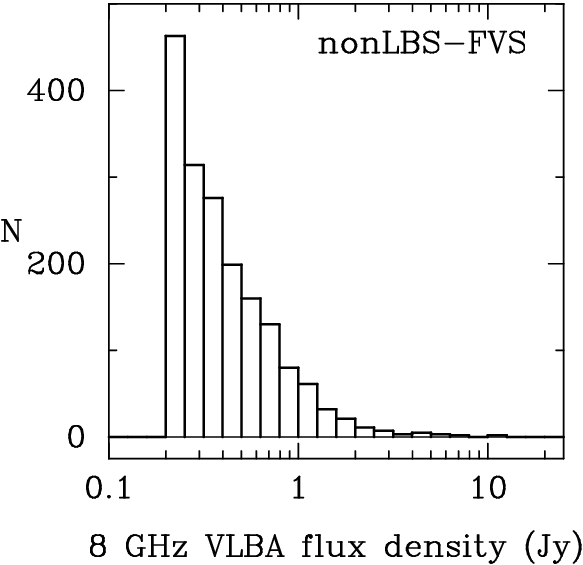}
}
\end{center}
\caption{\label{f:flux_VCSX}
Distribution of parsec-scale flux density measured by VLBA at 8 GHz for
the complete flux-density-limited sample (FVS) with
$S_\mathrm{VLBI}>0.2$~Jy and declination $\delta>-30^\circ$. {\it Top
panel:} Positive {\it Fermi}-VLBI associations. {\it Bottom panel:} {\it
Fermi}-VLBI non-associations. 
}
\end{figure}

An even stronger test is done by analyzing the full FVS sample of 1848
sources which is complete down to 0.2~Jy, separating them into LAT
detected and LAT non-detected (Figure~\ref{f:flux_VCSX}). We assume that
LAT non-detected objects have statistically lower $\gamma$-ray photon flux and
therefore should have lower radio flux. The Kolmogorov-Smirnov test
shows with a confidence greater that 99.99~per~cent that the
distributions of the parsec-scale flux density for the LAT detected and
non-detected VLBI sources are drawn from different parent 
distributions. The median flux densities of these two distributions differ
by a factor of 2.5: 0.84~Jy versus 0.34~Jy. This strongly supports
the correlation between $\gamma$-ray and parsec-scale radio emission,
confirming again the recent finding by \cite{FM2}.

\section{Summary}
\label{s:summary}

It has been shown that VLBI provides a very efficient tool to identify
bright $\gamma$-ray detections which have poor positional accuracy.
Application of this method is especially important for \Fermi LAT
identifications near the Galactic plane where VLBI observations are
affected by absorption and extended Galactic emission to a much lesser
degree than other radio observations.
It is suggested that this method be incorporated into the process of
identification of the \Fermi LAT catalogs and for estimation of
systematics in $\gamma$-ray positions. The results of $\gamma$-ray
source identification by \cite{LATBSL} have been confirmed. Six new
identifications are reported for a follow~up analysis, all within
$10^\circ$ from the Galactic plane.
It is estimated that more than 95~per~cent of the VLBI associations
found are firm identifications.

Direct cross-correlation of the \Fermi LAT and VLBI catalogs has
confirmed with a very high level of significance the early finding of
EGRET \citep{3EG} and \Fermi \citep{LBAS} that most of the bright
$\gamma$-ray sources on the sky are compact blazars. It was found that
\Fermi LAT preferentially ``selects'' the brightest objects from a
flux-density-limited sample of radio loud parsec-scale jets which
supports the hypothesis of a direct correlation and physical connection
between emission in the $\gamma$-ray and radio bands.


\acknowledgments 
The author thanks R.~Porcas and the anonymous referee for thorough
reading of the manuscript and useful comments.
The authors wish to thank L.~Petrov, the MOJAVE team, and the \Fermi LAT
team for useful discussions.
The author also thanks MPIfR PhD student Kirill Sokolovsky for help.
The author thanks his family, G.~Lipunova, A.~Kovaleva, and M.~Kovalev
for support.
Part of this project was done while the author was working as a research
fellow of the Alexander von~Humboldt Foundation. The author was partly
supported by the Russian Foundation for Basic Research (project
08-02-00545).
%
%
This research has made use of the NASA/IPAC Extragalactic Database (NED)
which is operated by the Jet Propulsion Laboratory, California Institute
of Technology, under contract with the National Aeronautics and Space
Administration. This research has made use of NASA's Astrophysics Data
System.

{\it Facilities:} \facility{VLBA, LBA, \Fermi (LAT)}.


\end{document}